\documentclass[conference]{IEEEtran}
\IEEEoverridecommandlockouts
\usepackage{cite}
\usepackage{amsmath,amssymb,amsfonts}
\usepackage{algorithmic}
\usepackage{graphicx}
\usepackage{textcomp}
\usepackage{xcolor}
\usepackage{amsmath}
\usepackage{xspace}
\usepackage{lettrine}

\def\BibTeX{{\rm B\kern-.05em{\sc i\kern-.025em b}\kern-.08em
    T\kern-.1667em\lower.7ex\hbox{E}\kern-.125emX}}
\begin{document}
\title{Flexible Beamforming in B5G for Improving  Tethered UAV Coverage over Smart Environments 
{\footnotesize \textsuperscript{}}
\thanks{}
}
\author{\IEEEauthorblockN{1\textsuperscript{st} Abdu Saif}
\IEEEauthorblockA{\textit{Faculty of Engineering\& IT} \\
\textit{Taiz University }\\
Taiz, Yemen\\
abdu.saif@taiz.edu.ye}
\and
\IEEEauthorblockN{2\textsuperscript{nd} Nor~Shahida~Mohd~Shah}
\IEEEauthorblockA{\textit{Faculty of Engineering Technology,}\\
\textit{Universiti Tun Hussein Onn Malaysia}\\
Johor, Malaysia\\
shahida@uthm.edu.my}

\and
\IEEEauthorblockN{3\textsuperscript{rd} Soreen Ameen Fattah}
\IEEEauthorblockA{\textit{College of Medicine, University of Babylon}\\
Babylon, Iraq\\
soreen.fattah@uobabylon.edu.iq}

\and
\IEEEauthorblockN{4\textsuperscript{th} Saeed Hamood Alsamhi}
\IEEEauthorblockA{\textit{Insight Centre for Data Analytics}\\
\textit{University of Galway}\\
Galway, Ireland\\
saeed.alsamhi@insight-centre.org
}
\and
\IEEEauthorblockN{5\textsuperscript{th} Santosh Kumar}
\IEEEauthorblockA{\textit{Department of CSE, IIIT Naya Raipur}\\
Chhattisgarh, India\\
santosh.rs.cse12@iitbhu.ac.in}
\and
\IEEEauthorblockN{6\textsuperscript{th} Ali Saad Al khuraib}
\IEEEauthorblockA{\textit{Information science and technology }\\
UKM, Riyadh, Malaysia\\
alipy60@yahoo.com}}
\maketitle
\begin{abstract}
Unmanned Aerial Vehicles (UAVs) are being used for wireless communications in smart environments. However, the need for mobility, scalability of data transmission over wide areas, and the required coverage area make UAV beamforming essential for better coverage and user experience. To this end, we propose a flexible beamforming approach to improve tethered UAV coverage quality and maximize the number of users experiencing the minimum required rate in any target environment. Our solution demonstrates a significant achievement in flexible beamforming in smart environments, including urban, suburban, dense, and high-rise urban. Furthermore, the beamforming gain is mainly concentrated in the target to improve the coverage area based on various scenarios. Simulation results show that the proposed approach can achieve a significantly received flexible power beam that focuses the transmitted signal towards the receiver and improves received power by reducing signal power spread. In the case of no beamforming, signal power spreads out as distance increases, reducing the signal strength. Furthermore, our proposed solution is suitable for improving UAV coverage and reliability in smart and harsh environments.
\end{abstract}
\section{Introduction}
\lettrine{U}{}nmanned aerial vehicles (UAVs) have been increasingly used to boost the system performance of terrestrial networks with device-to-device (D2D) communication, particularly in smart environments and harsh environments where infrastructure network failure occurs. However, to establish the viability of remote surveillance, monitoring, relief operations, package delivery, and communication backhaul infrastructure, there is a need for accurate air-to-ground (AG) propagation channel models to design and evaluate UAVs' communication links for both control and non-payload data transmissions \cite{rr1}. 

The main challenge in disaster-affected areas is the physical damage to network infrastructures or power outages. In such scenarios, UAVs act as flying base stations to satisfy the communication requirements of people in the affected areas \cite{saif2022uav,sa1}. UAVs can also enable the Internet of Things (IoT) in remote environments without traditional communication infrastructure such as smart city~\cite{sa2,sa4}. However, researchers face challenges such as the lack of backhaul connectivity, the need for D2D communication, the mobility of underlay devices, or the unavailability of overlay devices due to failure or adversarial attacks~\cite{rr3}. In this context, beamforming techniques can improve the coverage and reliability of UAV-based communication systems in smart environments. 

Optimizing the bandwidth and power allocations of ground user devices' communications is a crucial challenge for UAV-based communication systems. In this regard, we propose a novel UAV-Direct approach, where a UAV can assist in resource allocation management to improve the system throughput \cite{r2}. The TUAV can also act as a relay to maintain communication links between devices that are out of range of each other\cite{saif2021energy}. A dynamic game and power control among UAVs is proposed to address the problem of interference-aware path planning for a network of cellular-connected UAVs\cite{saif2021efficient}. Each UAV acts as cellular user equipment (UE) in this game and aims to balance energy efficiency, wireless latency, and interference caused on the ground network.  In \cite{r3}, the proposed algorithm achieves a sub-game perfect Nash equilibrium upon convergence, leading to better wireless latency per UAV and rate per ground UE. Simulation results demonstrate the proposed scheme's effectiveness, which can improve the coverage and reliability of UAV-based communication systems in smart environments.

UAV-based communication systems can be affected by various factors, such as interference, obstacles, and limited communication infrastructure. One solution to these challenges is beamforming, a technique that can improve the quality and range of wireless signals by focusing the transmission in a specific direction. By beamforming, UAVs can transmit data more efficiently, even in challenging communication environments, which can help improve the overall effectiveness of these systems. Various beamforming techniques, such as phased array and adaptive beamforming, can be used in UAV-based communication systems\cite{ab2,ab3}. Phased array beamforming involves using multiple antennas to steer the transmission beam in a specific direction, while adaptive beamforming uses algorithms to dynamically adjust the direction of the beam based on the changing communication environment. Both techniques can effectively improve the coverage and reliability of UAV-based communication systems.
In this paper, we propose a flexible beamforming approach to improve Tethered UAV coverage quality and maximize the number of users experiencing a minimum required rate in any target environment while demonstrating significant achievement in various smart environments, including urban, sub-urban, dense, and high-rise urban scenarios. Our solution is suitable for improving tethered UAV coverage and reliability in smart environments. The proposed solution of flexible beamforming in B5G for improving tethered UAV coverage over smart environments is motivated by the need to improve coverage and user experience in wireless communications and the challenge of achieving mobility and scalability of data transmission over wide areas. The potential of tethered UAVs to address the challenges of mobility and scalability and improve coverage quality and the user experience in smart environments is also a key motivation.
Ultimately, the proposed solution aims to provide a suitable solution to improve UAV coverage and reliability in both smart and harsh environments, including public safety, disaster response, and data collection in smart cities.

The contributions are summarized as follows: 
\begin{itemize}   
\item We introduce  the importance of beamforming for improving Tethered UAV coverage in smart environments.
\item We introduce the concept of flexible beamforming to address the challenges of mobility and scalability of data transmission over smart environments. Furthermore, we highlight the potential of tethered UAVs for improving coverage quality and the user experience in smart environments.
\end{itemize}
\section{System model}
The proposed system model for the flexible beamforming approach in tethered UAV coverage consists of a ground control station (GCS) and a tethered UAV equipped with multiple antennas. The tethered UAV hovers at a fixed height and is connected to the ground station via a tether, providing power to the UAV. The smart environment is divided into multiple grid cells. Each grid cell contains one or more user equipment (UE) devices, the intended receivers of the UAV's wireless signals. 
The proposed approach uses a flexible beamforming technique that adapts the direction of the UAV's wireless signal to maximize the number of UE devices that receive the minimum required data rate. Furthermore, the approach optimizes the beamforming weights based on the channel state information (CSI) between the UAV and each UE device. The optimization problem aims to maximize the number of UE devices that achieve a minimum data rate threshold while limiting the total power consumption of the UAV.
Therefore, it dynamically adjusts the beamforming direction and weights as the UAV moves, and the channel conditions change due to the movement of UE devices or other environmental factors.

\begin{figure}[htbp]
\includegraphics[width=1\linewidth]{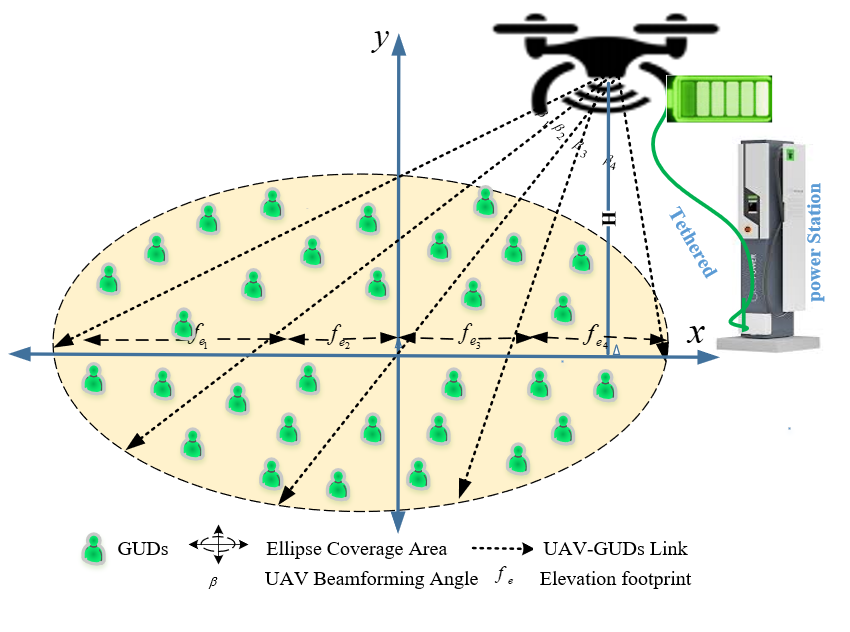}
\caption{System Model }
\label{f1}
\end{figure}

Fig.\ref{f1} show the scenario of the proposed tethered UAV  with adjust beamforming waves for improving Tethered UAV coverage in smart environments. we assuming the UAV coverage modeled as an ellipse equation as follow: 
\begin{equation}
\begin{split}
 {\frac {{x_{{i}}}^{2}}{{a_{{i}}}^{2}}}+{\frac {{y_{{i}}}^{2}}{{b_{{i}}
}^{2}}}=1 
\Rightarrow  x_{{i}}=\pm \frac {\sqrt {{b_{{i}}}^{2}-{y_{{i}}}^{2}}a_{{i}}}{b_{{i}}}, 
y_{{i}}=\pm {\frac {\sqrt {{a_{{i}}}^{2}-{x_{{i}}}^{2}}b_{{i}}}{a_{{i}}}}
\end{split}
\end{equation}
 The distance between the UAV and the user devices in the ellipse coverage area 
\begin{equation}
d=\sqrt { \left( x_{{j}}-a_{{i}}\sqrt {1-{\frac {{y_{{i}}}^{2}}{{b_{{i
}}}^{2}}}} \right) ^{2}+ \left( y_{{j}}-b_{{i}}\sqrt {1-{\frac {{x_{i}}^{2}}{{a_{i}}^{2}}}} \right) ^{2}+{z_{j}}^{2}}
\label{Eq1}
\end{equation}
where $x_{i}$ and $y_{i}$ are the coordinates of a point on the ellipse
$a$, $b$ are the length of the major axis of the ellipse, to the maximum width and height of the coverage area.
In this context, any point  $\left(x_{j},y_{j}\right)$ that satisfies the equation lies on the ellipse, which represents the boundary of the coverage area of the UAV. Hence, the  enhance the UAV coverage based on adjust  beamforming  at the desired location.
\begin{figure}[htbp]
\includegraphics[width=1\linewidth]{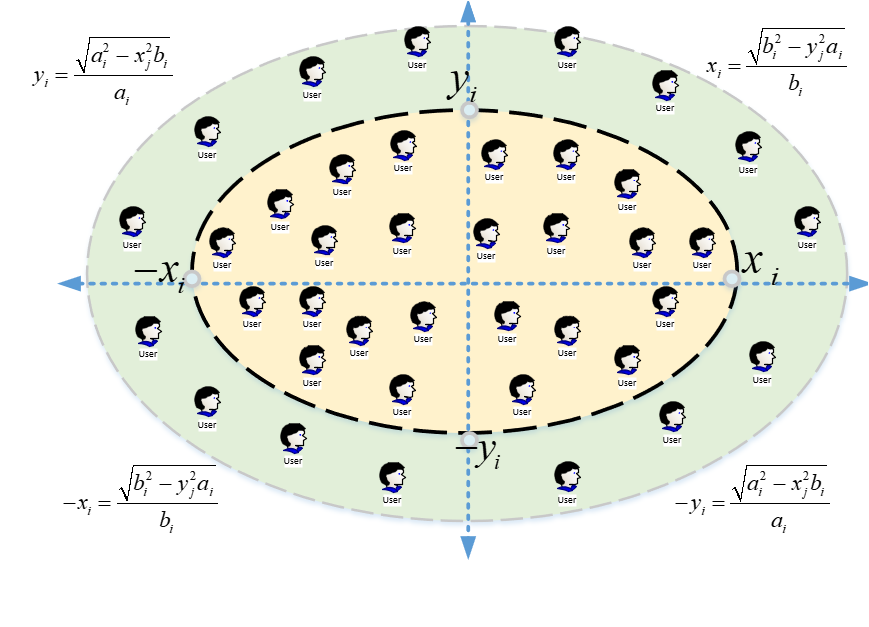}
\caption{TUAV Coverage range }
\label{f2}
\end{figure}
\subsection{Air-to-Ground downlink Channel}
As shown in the Fig.2 , the optimal coverage area of a UAV range is determined by several factors, including altitude, sensor capabilities, and mission requirements. Flying at a higher altitude can increase the coverage area, but it may reduce the resolution of the sensors. 
The sensors on board the UAV can also affect the coverage area by determining how far and how well targets can be detected. The coverage area ultimately depends on the specific mission requirements, which may require a larger or smaller coverage area depending on the need for high-resolution imagery or a broad overview of a large area.
Therefore, the channel between UAV  and ground user devices in downlink are characterized as A-G channel links. 
The gain of channel link from UAV to the GUDs is given by \cite{r8}.
\begin{equation}
h=\left\{\begin{matrix}
(\sqrt{h^{2} + x_{i}^{2}+ y_{i}^{2}})^{-\alpha} \  \ \text{for} \ \text{LoS} \\
\eta(\sqrt{h^{2} + x{i}^{2}+ y_{i}^{2}})^{-\alpha} \  \ \text{for} \ \text{NLoS} 
\end{matrix}\right.
\label{equ10}
\end{equation}
where, $\left( x_{i}, y_{i} \right)$ is a 2D coordinate system of the ground user devices  and , $\alpha$ is path-loss exponent associated with the A-G link for ground user devices  and $\eta$ is  excess-loss encountered for NLoS link from UAVs and ground user devices. 
The probability of Line of sight (LoS) link represent as  a function  of ground user devices elevation angle $\theta$ and environment parameters $a$, $b$ such that  LoS probability associated with  ground user devices  can be shown as follows \cite{r3}.
\begin{equation}
P_{LoS} =\frac{1}{1+a\cdot\mbox{exp}(-b_(\theta -a))},
\label{equ11}
\end{equation}
where, $a$ and $b$ are parameters that associated to the S-curve that  vary according to the environment, i.e., sub-urban, urban, dense urban and high rise urban. The elevation angle $\theta$ of the ground user devices , in (\ref{equ11}), is measured in degrees and is given as follows.
\begin{equation}
\theta = \frac{180}{\pi} \mbox{sin}^{-1}\left(\frac{h}{\sqrt{x_{i}^2+y_{i}^2+h^2}}\right).
\label{12}
\end{equation}
The NLoS probability of the ground user devices  can be obtained as $P_{\text{NLoS}} = 1-P_{\text{LoS}}$.
The average channel gain for link between UAVs and ground user devices  is denoted  as follows.
\begin{equation}
\begin{split}
\bar{h} = P_{LoS}\left(\sqrt{h^2 + x_{i}^2+ y_{i}^2}\right)^{-\alpha} \\
+ P_{NLoS}\eta\left(\sqrt{h^{2} + x_{i}^{2}+ y_{i}^{2}}\right)^{-\alpha}.
\end{split}
\label{13} 
\end{equation}
\subsection{UAV Beam-forming  analysis }
Beamforming is a technique used to focus the antenna radiation pattern in a specific direction or region, which helps to improve the received signal quality at the receiver end. Beamforming in UAVs can significantly improve coverage and user experience, particularly in smart environments  where communication infrastructure may be unavailable or damaged.
Several factors need to be considered in the analysis of UAV beamforming, including the physical environment, the number and position of UAVs, the number of users, and the desired communication quality. Additionally, the beamforming algorithm and optimization strategy play a crucial role in determining the effectiveness of the beamforming technique in the given scenario. UAV beamforming is to use a flexible beamforming method, which can adapt to changes in the environment and the number of users. This approach can maximize the number of users experiencing a minimum required rate in any target environment while also ensuring that the beamforming gain is mainly concentrated in the target coverage area. Furthermore , As shown in Fig.3, UAV beamforming is a technique that directs signals from a UAV's antenna array to a specific location or target. It adjusts the phase and amplitude of signals to control the direction and shape of multiple beams to maximize signal strength. Ground nodes' elevation angle is a crucial factor that determines the direction and shape of beams.
\begin{figure*}[t!]
    \centering
    \includegraphics[width=1\linewidth]{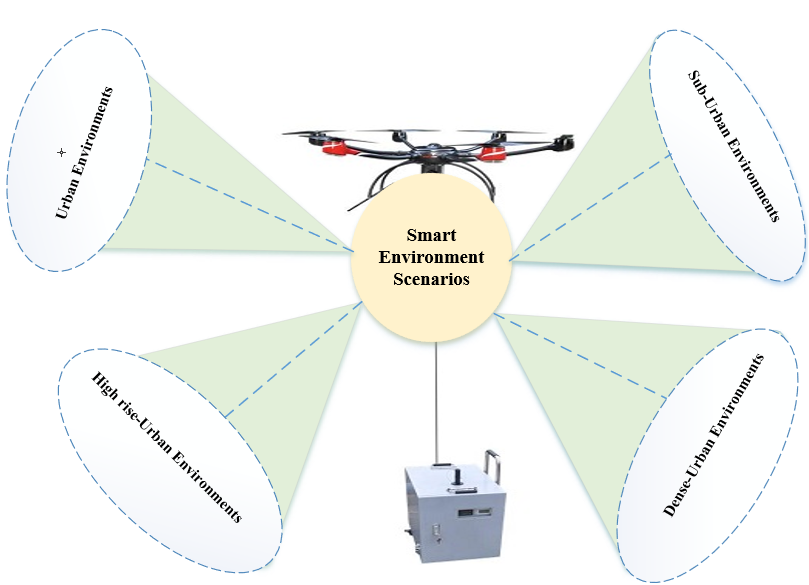}
    \caption{TUAV beamforming over smart environments scenarios}
    \label{f3}
\end{figure*}
The desing of the beamforming over the line of sight channel (LoS) aim of maximizing the number of user devices that received the coverage service from the UAV. Let $w(\phi)=\alpha(\phi)$ represent the beamforming vector (transmitted beam by the UAV) and the term $\left| \alpha (\theta)^{H} {w}(\phi) \right|^2$ be the array gain which can be computed as~\cite{r9}
\begin{equation*} \Big |\boldsymbol {a}(\theta)^{H}\boldsymbol {w}(\phi)\Big |^{2}=\Bigg |\frac {\sin {\left({\frac {M\pi }{2}(\sin {\theta }-\sin {\phi })}\right)}}{M\sin {\left({\frac {\pi }{2}(\sin {\theta }-\sin {\phi })}\right)}} \Bigg |^{2}.\tag{7}\end{equation*}
where, the beam angle $\phi$ describes a beamforming vector, $\theta$ is the user devices elevation angle and $M$ an  element antenna array. 
In addition, the beamforming of the UAV antenna illustrates the azimuth footprint, and the elevation beam width will narrow as the vehicles approach the target line. The value of the elevation beam width determines the current elevation footprint for a given effective antenna height and current distance. 
\begin{equation}
    fe_{k}=\frac{(h^{2}_{n}+r^{2}_{k})tan(\beta_{k})}{h_{n}+r_{k}tan(\beta_{k})}
    \label{equ8}
    \tag{8}\
\end{equation} 
where , r is the maximum boundary horizontal distance of the targeted area to acreage.
\begin{table}[htbp]
\caption{}
\begin{center}
\small
\begin{tabular}{|p{2cm}|p{1.8cm}|p{1.8cm}|p{1.8cm}|}
\hline
\textbf{Table}&\multicolumn{3}{|c|}{\textbf{Simulation Parameters}} \\
\cline{2-4} 
\textbf{Items}& \textbf{\textit{values}}& \textbf{\textit{Items}}& \textbf{\textit{values}} \\
\hline
Bandwidth& $B=10$~MHz& Noise figure (dB) & $nf=5$ \\	 \hline
Carrier frequency&$f=2.4$~GHz&UAV TX power & $20$~dBm \\ \hline
Transmit antenna gain(dBi)&$gt = 10$&Receive antenna gain(dBi)& $gr = 10$ \\ \hline
Urban factors&$a = 9.61$, $b=0.16$,&$\eta_{\mbox{LoS}} = 1$& $\eta_{\mbox{NLoS}} = 20$ \\ \hline
Sub-urban factors & $a=4.88$, $b=0.43$,&$\eta_{\mbox{LoS}}=1$ &$\mu_{\mbox{NLoS}}=21$\\ \hline
Dense-Urban & $a = 12.08$, $b=0.11$ &$\eta_{\mbox{LoS}} = 1.6$&$\eta_{\mbox{NLoS}}= 23$
 \\ \hline
 High-rise Urban & $a=15.05$, $b=0.08$, &$\eta_{\mbox{LoS}}=2.3$,& $\eta_{\mbox{NLoS}}=34$ \\ \hline
\hline
\multicolumn{4}{l}{$^{\mathrm{}}$}
\end{tabular}
\label{tab1}
\end{center}
\end{table}
\section{Simulation Results and Discussion}
\begin{figure}[t!]
	\includegraphics[width=1\linewidth]{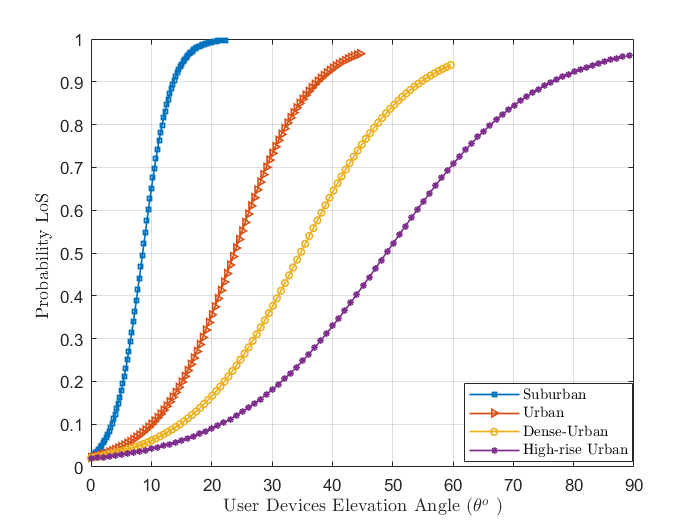}
	\caption{Probability of Los versus GUDs elevation Angle with multi environments }
	\label{f4}
\end{figure}

\begin{figure}[t!]
	\includegraphics[width=1\linewidth]{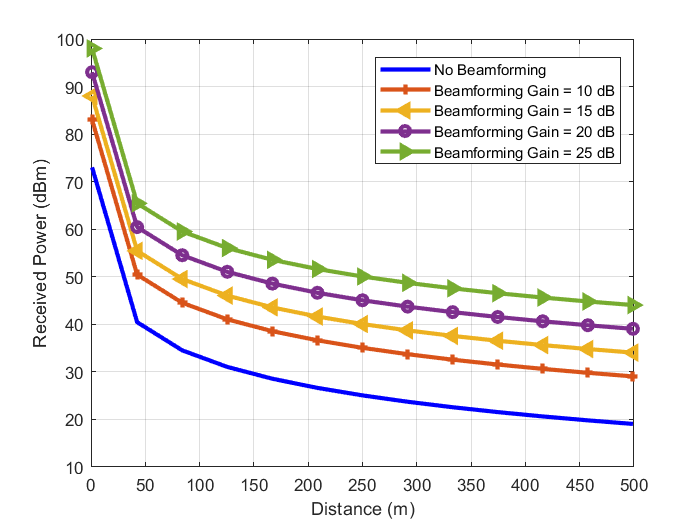}
	\caption{Performance of UAV Beamforming versus Distance in urban areas}
	\label{f5}
\end{figure}

\begin{figure}[t!]
   \includegraphics[width=1\linewidth]{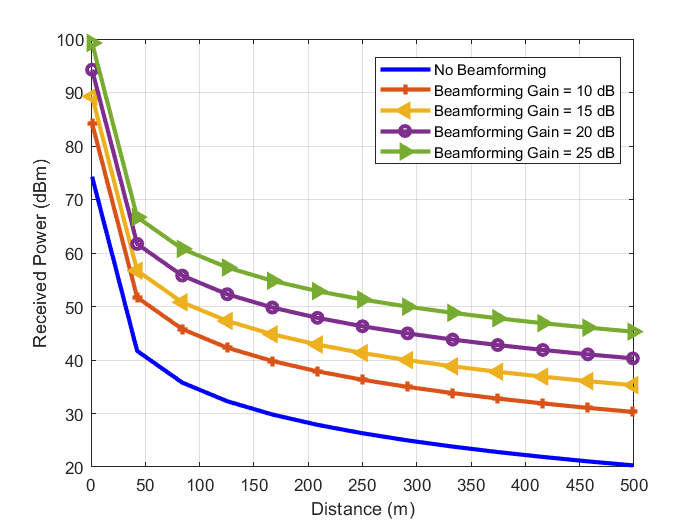}
	\caption{Performance of UAV Beamforming versus Distance in suburban areas}
	\label{f6}
\end{figure}

\begin{figure}[t!]
   \includegraphics[width=1\linewidth]{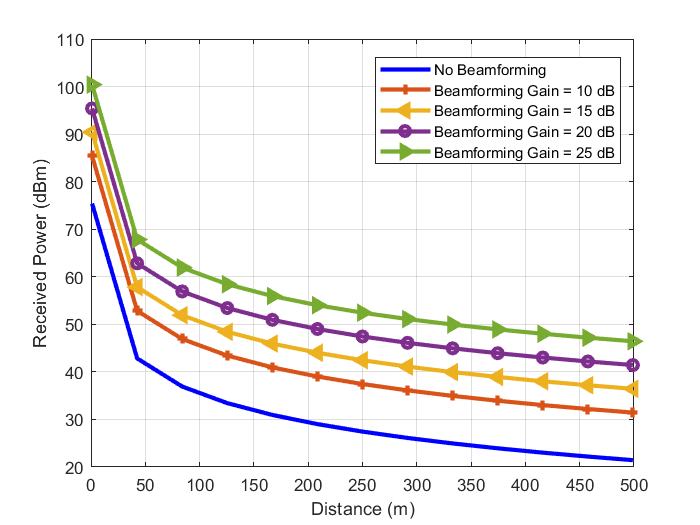}
	\caption{Performance of UAV Beamforming versus Distance in Dense urban areas}
	\label{f7}
\end{figure}

\begin{figure}[t!]
   \includegraphics[width=1\linewidth]{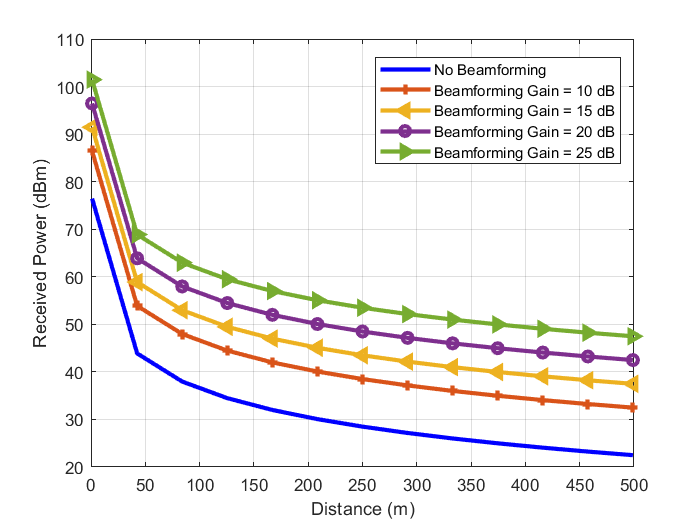}
	\caption{Performance of UAV Beamforming versus Distance in high rise urban areas}
	\label{f8}
\end{figure}

In this section, we present simulation results to demonstrate the performance of the proposed schemes. The results evaluate the Line of Sight (LoS) probability, received power for the UAV system at various elevation angles, and Ground User Device (GUD) distances. The simulations use carrier frequencies of 2.4 GHz and assume that user devices are randomly distributed in each communication scenario. The transmission distance between source and destination user devices is kept at 500 m for each scenario, while the elevation angle for user devices is maintained between 0$^\circ$ and 90$^\circ$.
The \ref{f4} the performance probability of line of sight (PLoS) by adjusting the UAV beamforming in various environments, based on the coverage distance. The results show that the achieved PLoS increases with higher elevation angles for all environments, as the UAV beamforming adjusts to the specific area. However, the performance of PLoS varies considerably across different environments  based on the UAV beamforming  adjustable to the targeted  area  with versus higher elevation angles.
In the case of  high-rise urban environment provides the best SINR performance for the ground nodes , while the urban environment provides the worst performance. This is likely due to the high interference and blockage effects in urban environments, which make it challenging for the UAV beamforming to direct the signal towards the GUDs effectively. Furthermore,  figure of results demonstrates that UAV beamforming can be an effective technique to improve the signal quality for GUDs in different environments, but its performance can vary significantly depending on the specific environment.
In the context of a tethered UAV receiving power in different smart environments scenarios, the  beamforming performance shows the received power level of the UAV in different environments including urban, sub-urban, dense urban, and high-rise urban. The  Fig.~\ref{f5} and \ref{f6} provide an analysis of the performance of received power (dB) versus distance for different environment scenarios such as urban and sub-urban. The analysis shows that signal strength decreases as the distance from the source increases, but beamforming reduce loss of signal strength by increasing the gain. Moreover, the rate of signal strength decrease varies depending on the environment. In urban environments with many buildings and obstacles, the path loss increases more quickly than in suburban or rural areas. The reasons include: urban environments have a high density of buildings and other structures, which results in high levels of interference and signal attenuation; suburban areas contain a significant number of structures that affect signal strength; dense urban environments are similar to urban environments, but with an even higher density of structures; and high-rise urban environments are characterized by tall buildings that block or reflect wireless signals. 
The beamforming performance figure shows how the tethered UAV's received power level varies in each of these different environments. For example, the received power level is typically highest in sub-urban environments and lowest in high-rise urban environments. In addition, the figure shows that beamforming performance is improved in environments with fewer obstructions and less interference. Briefly, the figures of beamforming performance for a tethered UAV that received power over different smart environment scenarios help identify which environments are most suitable for beamforming technology and how the technology is optimized to perform well in different environments. Furthermore, Without beamforming, signal power spreads out as distance increases, reducing signal strength. Adjustable beamforming can focus the transmitted signal towards the receiver and improve received power by reducing signal power spread.
Fig. \ref{f7} and \ref{f8} illustrate the performance of received power versus distance in dense urban and high-rise urban environments. Received power is a measure of the probability that a wireless signal will be successfully received by a receiver within a given area. The results  demonstrate that the received power decreases with distance, with and without beamforming gain.
Furthermore, comparing the received power results in dense and high-rise urban environments provides valuable insight into the trade-offs between coverage and network density in different types of environments. For instance, in dense urban environments with high-rise buildings, the received power may be lower due to interference and signal blockage, but a higher density of network infrastructure can help improve coverage. In contrast, in suburban or rural areas with fewer obstacles, a lower density of network infrastructure may still provide adequate coverage.
The results of the analysis can inform decision-making for network planning and deployment in different environments, helping network designers optimize system performance, reliability, and coverage.
\section{ conclusion}
The proposed solution of flexible beamforming in B5G for improving tethered UAV coverage over smart environments offers a promising solution to address the challenges of achieving mobility and scalability of data transmission over wide areas using UAVs while improving coverage quality and user experience in smart environments. By providing a flexible beamforming approach that maximizes the number of users experiencing a minimum required rate in any target environment, the proposed solution demonstrates a significant achievement in various smart environments, including urban, sub-urban, dense urban, and high-rise urban scenarios. Furthermore, by concentrating the beamforming gain mainly in the target area, the proposed approach aims to improve the coverage area and achieve significant received power in the target area. Ultimately, the proposed solution offers a suitable way to improve UAV coverage and reliability in both smart and harsh environments, which could have significant implications for various industries and applications.
\section{Acknowledgment}
This research is supported by Universiti Tun Hussein Onn Malaysia (UTHM) through Tier 1 (vot Q444)
\bibliography{Referencesbib.bib}
\bibliographystyle{IEEEtran}
\vspace{12pt}
\end{document}